\documentclass[preprint,prd,noshowpacs,nofootinbib]{revtex4}

\usepackage{color} 
     
\usepackage{amssymb}
\usepackage{amsmath}
\usepackage{graphicx}
     
\usepackage{pdfpages}

\usepackage{epstopdf}

\usepackage{hyperref}

\begin{document}

\title{Modified commutation relationships from the Berry-Keating program}

\author{Michael Bishop}
\email{mibishop@mail.fresnostate.edu}

\affiliation{Mathematics Department, California State University Fresno, Fresno, CA 93740 USA}
\author{Erick Aiken}
\email{erickaik@mail.fresnostate.edu}

\affiliation{Physics Department, California State University Fresno, Fresno, CA 93740 USA}

\author{Douglas Singleton}
\email{dougs@csufresno.edu}

\affiliation{Physics Department, California State University Fresno, Fresno, CA 93740 USA}

\date{\today}

\begin{abstract}
Current approaches to quantum gravity suggest there should be a modification of the standard quantum mechanical commutator, $[{\hat x} , {\hat p}] = i \hbar$. 
Typical modifications are phenomenological and designed to result in a minimal length scale. 
As a motivating principle for the modification of the position and momentum commutator, we assume the validity of a version of the Bender-Brody-M\"uller variant of the Berry-Keating approach to the Riemann hypothesis.
We arrive at a family of modified position and momentum operators, and their associated modified commutator, which lead to a minimal length scale.    
Additionally, this larger family generalizes the Bender-Brody-M\"uller approach to the Riemann hypothesis.

\end{abstract}

\maketitle

\section{Introduction}

At present there is no agreed upon approach to quantizing gravity. 
However, there are general arguments that no 
matter what final form quantum gravity takes, it should have some 
non-zero minimal distance scale $\Delta x_0$. 
String theory based arguments lead to such a minimum absolute length 
scale (see \cite{garay-1995} and the references 
therein for a survey). 
Many works \cite{kempf-1992} \cite{kempf-1993} \cite{kempf-1994} 
\cite{KMM-1994} \cite{adler-1999} \cite{das1} \cite{das2} have shown how  
a modification of the standard quantum commutator, $[{\hat x}, {\hat
p}] = i$ leads to a minimal length scale
(in this paper we choose units so that $\hbar =1$). There are also works 
which propose a minimal length scale by modifying the 
standard spatial and/or momentum commutators by allowing 
$[{\hat x_i}, {\hat x_j}]$ and/or 
$[{\hat p_i}, {\hat p_j}]$ to be non-zero \cite{piero-2007}. The two
approaches of modifying either $[{\hat x}, {\hat 
p}]$ or $[{\hat x_i}, {\hat x_j}]$ and/or 
$[{\hat p_i}, {\hat p_j}]$ are related. A nice and current
overview of minimal length scales arising from quantum gravity is \cite{hossenfelder}.
In this work we will focus on the introduction of a minimal length 
scale via a modification of $[{\hat x}, {\hat p}] = i$. 

In \cite{KMM-1994}, a simple modification of the quantum 
commutation relationship between the position operator (${\hat x}$) 
and momentum operator (${\hat p}$) was proposed of the form
\begin{equation}
\label{KMM-CR}
[{\hat x} , {\hat p}] = i (1 + \beta {\hat p}^2) ~,
\end{equation}
\noindent where $\beta$ is an arbitrary parameter which is assumed to 
come  from quantum gravity. Using \eqref{KMM-CR} and the standard 
relationship between the quantum commutators and uncertainties gave

\begin{equation}
\label{KMM-UP}
\Delta x \Delta p \ge \frac{1}{2} \left( 1 + \beta \Delta p ^2 + \beta 
\langle \hat p \rangle ^2 \right) ~,
\end{equation}


\noindent which in turn gave a minimal distance of $\Delta x _0 = 
\sqrt{\beta}$. 
One criticism of this approach is 
that it is purely phenomenological, bottom-up. 
The parameter $\beta$ is not determined, and even the specific form of the 
modified commutation relationship in \eqref{KMM-CR} is an 
assumption. 
Having an undetermined parameter such as 
$\beta$ is similar to the introduction of the reduced 
Planck's constant, $\hbar$, which was originally   
introduced as a parameter to fit the observed blackbody spectrum. 

There are various physical motivations which support the modified 
commutator and uncertainty relationship like those given in equations
\eqref{KMM-CR} and \eqref{KMM-UP}. 
In reference \cite{maggiore} arguments are made that at low energy the uncertainty relationship
is dominated by the Compton length of an object which leads to the 
usual relationship $\Delta x \sim \frac{1}{\Delta p}$. 
At high energy the uncertainty relationship is dominated by the Schwarzschild radius
of the object which leads to relationship $\Delta x \sim \Delta p$.
Combining these regimes linearly leads to a modified uncertainty
relationship similar to \eqref{KMM-UP}. There are also string
theory arguments \cite{amati, amati2, gross} based on looking
at colliding strings in the eikonal limit. 
The parameter $\beta$ is found to be related to the Planck scale in all of these approaches \cite{maggiore, amati, amati2, gross}. 

We propose an approach to obtaining modified commutation 
relationships which we consider to be between the 
phenomenological approach of \cite{KMM-1994} and the physical approach
of \cite{maggiore, amati, amati2, gross}. 
Our method is motivated by the Bender-Brody-M\"uller approach 
\cite{BBM-2017} to the Riemann hypothesis  \cite{Riemann-1859}. 
The Riemann hypothesis deals with the non-trivial zeros of the Riemann 
zeta function and is connected with the distribution of prime numbers.
The Riemann zeta function is given by

\begin{equation}
\label{R-zeta}
\zeta (z) = \sum _{n=1} ^\infty \frac{1}{n^z} = \frac{1}{\Gamma (z)} 
\int _0 ^\infty \frac{t^{z-1}}{e^t -1} dt 
\end{equation}

\noindent where $\Gamma (z) = \int _0 ^\infty e^{-t} t^{z-1} dt$ is 
the usual gamma function. Using the 
integral expression in \eqref{R-zeta} one obtains a 
reflection formula for the Riemann zeta function 

\begin{equation}
\label{zeta-ref}
\zeta(z) = 2^z \pi ^{z-1} \sin (\pi z/2) \Gamma (1-z) \zeta (1-z).
\end{equation}

From \eqref{zeta-ref} one can see that the Riemann zeta function has
{\it trivial} zeros at the negative even integers, $z=-2n$ due to the $\sin (\pi z/2)$ term. 
Riemann noticed that there were also non-trivial zeros which occurred along the line 
$Re (z) =\frac{1}{2}$. 
Specifically there were zeroes at the complex values $z_n = \frac{1}{2} + i t_n$ where 
$n=1, 2, 3 ....$ and  $t_1 =  14.135$ , $t_2 =21.022$, $t_3 = 25.011$ 
{\it etc.}  \footnote{The first hundred non-trivial zeros can be found at 
${\rm http://www.dtc.umn.edu/~odlyzko/zeta\_tables/zeros2}$.} 
The Riemann hypothesis states that all of these nontrivial zeros lie on this 
line $z= \frac{1} {2} + i t$.

From the discrete nature of the imaginary part of the non-trivial 
zeros of the Riemann zeta function, it was conjectured that 
these non-trivial zeros were related to an eigenvalue problem.
The general suggestion is there exists some operator, ${\hat H}$, 
whose eigenvalues are the imaginary parts of the non-trivial 
zeros of the Riemann zeta function. 
This is called the Hilbert-Polya conjecture. 
The operator ${\hat H}$ is conventionally called the ``Hamiltonian", although it is not connected with 
the energy of any system, and does not even have the dimensions of energy. 
Berry \cite{berry-1986} and Keating  \cite{berry-1999} suggested a version of this proposal
where the quantum version of the operator ${\hat H}$ should reduce to the classical operator $H=xp$. 
One proposal is to take ${\hat H} = \frac{1}{2}({\hat x} {\hat p} + {\hat p} {\hat x})$ thus taking into account that the order of the quantum operators ${\hat x}$ and ${\hat p}$ matters. 
This form of ${\hat H}$ is proportional to the one dimensional virial operator\footnote{ The virial 
operator is $A= \frac{1}{2} ({\hat x} {\hat p} + {\hat p} {\hat x})$, in terms of which the dilation
transformation is $D = e^{iA\theta}$ with $\theta$ being some scaling parameter.} which is the generator
for scaling/dilation transformations. 
In a conformal theory with no length scale, the appearance of scaling/dilation transformations
makes sense since these transformations represent symmetries of such a theory. 
We argue for a minimal length scale which breaks scaling/dilation symmetry. 
Thus the appearance of a modified or broken dilation symmetry makes sense in the context of looking for a theory with a minimal, absolute length scale. 
Below we introduce an operator ${\hat \Delta}$ which modifies/breaks the virial operator and the related dilation symmetry. This leads to a minimum length.
In the discussion below we will drop the nomenclature of ``Hamiltonian" operator 
and simply say either operator or modified virial operator. 

Our proposal is to modify $\hat{x}$ and $\hat{p}$ so that they 
align with the recent attempt of Bender-Brody-M{\"u}ller 
to address the Riemann hypothesis through 
the Berry-Keating program. The modified operator proposed 
by Bender-Brody-M{\"u}ller \cite{BBM-2017} is

\begin{equation}
\label{BBM-H}
{\hat H} = {\hat \Delta_{BBM}}^{-1} ({\hat x} {\hat p} + {\hat p} {\hat x}) 
{\hat \Delta_{BBM}}  ~,
\end{equation}

\noindent where ${\hat \Delta_{BBM}} = 1-e^{-i {\hat p} \Delta x }$.  
When applied to an analytic function $f(x)$, this is the difference operator, 
${\hat \Delta_{BBM}}f(x)  = f(x) - f(x-\Delta x)$ between the values of the function at $x$ and $x-\Delta x$. 
In the original work of Bender-Brody-M{\"u}ller, $\Delta x$ equals $1$ so that the operator was the unit difference operator.
Here we retain the explicit distance scale through $\Delta x$ so we may discuss
modifications of the commutator that will lead to an explicit minimal distance. 
The modified operator in \eqref{BBM-H} is a combination of the virial operator
({\it i.e.} $({\hat x} {\hat p} + {\hat p} {\hat x})$) and the discrete difference operator
${\hat \Delta_{BBM}} = 1-e^{-i {\hat p} \Delta x}$ and its inverse. 
It is not clear how combining these two as in \eqref{BBM-H} would break conformal symmetry and lead to some absolute minimal length scale.  We find that the operator in \eqref{BBM-H} {\it does not}
work for the purpose of introducing a minimal distance. 
The way in which the operator in \eqref{BBM-H} fails leads us to a different form of the modified operator
\begin{equation}\label{ABSDelta}
{\hat \Delta_{ABS}} =  \frac{1}{2} (e^{kp} + e^{-kp})= \cosh (kp)~    
\end{equation}
where $k$ could be a real, imaginary, or complex constant. 
We will propose that a modified virial operator similar to that in \eqref{BBM-H} but with
${\hat \Delta}_{BBM}$ replaced by ${\hat \Delta}_{ABS}$ {\it which does} give a modified dilation symmetry
and introduces an absolute minimal length scale.  If a minimum length scale exists, functions should be approximately constant along any interval smaller than 
the minimum length scale. The operator $\hat{\Delta}_{ABS}$ takes the average of some function $f(x)$ with 
respect to some general shift of $\pm i k$, {\it i.e.} $\hat{\Delta}_{ABS} f(x) = \frac{1}{2}(f(x-ik)+ f(x+ik))$. If one takes $k= \pm i\frac{\Delta x}{2}$ then this equals $\frac{1}{2}(f(x-\frac{\Delta x}{2})+ f(x+\frac{\Delta x}{2}))$ which is an averaging of $f(x)$ over an interval $\Delta x$. This operator sends functions approximately to zero on intervals where they oscillate at length scales less than the minimal length scale, $\Delta x$.  We will find
that in order to get a modified commutator that leads to a minimal length, as well as modified position and momentum 
operators that have good limiting behavior, we need the $k$ in \eqref{ABSDelta} to be pure real. 
For a real $k$ this implies that the $\hat{\Delta}_{ABS}$ gives a shift of the function in the imaginary direction namely 
$\hat{\Delta}_{ABS} f(x) = \frac{1}{2}(f(x-ik)+ f(x+ik))$. 
In contrast the operator ${\hat \Delta}_{BBM}$ uses
a pure imaginary $k$ which gives a shift in the real direction. 

Returning to the operator in \eqref{BBM-H} and taking into account that ${\hat x}$ and ${\hat p}$ 
satisfy the standard commutator relation $[{\hat x} , {\hat p}] = i$, we can ``walk"
the operator ${\hat \Delta_{BBM}}$  through to the left and annihilate it with its 
inverse operator  ${\hat \Delta_{BBM}}^{-1}$. The operator in \eqref{BBM-H} becomes 

\begin{equation}
\label{BBM-H2}
{\hat H} = ({\hat x} {\hat p} + {\hat p} {\hat x}) - \frac{2 {\hat 
p} \Delta x e^{-i {\hat p} \Delta x}}{(1-e^{-i {\hat p} \Delta x})} ~.
\end{equation}

%

We use this modified virial operator to motivate a family of modified position and momentum operators and their modified commutator.  
The modified position and momentum operators, given below, are symmetric and lead to a minimum length scale similar to 
\cite{KMM-1994}. 
These modified position and momentum operators are also symmetric in an inner-product space which requires the wave function to decay exponentially in the large momentum limit. 
The modified virial operator, which we find below, gives a similar approach to the Riemann hypothesis as that suggested by 
Bender-Brody-M{\"u}ller \cite{BBM-2017}. 
There are important open questions \cite{bellissard} and additional discussion \cite{muller}
concerning the Bender-Brody-Mueller approach to the Riemann hypothesis. 
We do not resolve the major criticism of ``What is the Hilbert space used in the construction in reference 
\cite{BBM-2017}?". The family of operators we present may 
provide alternative avenues to resolving this criticism.  However, 
our main goal here is to use operators, such as given 
in equation \eqref{BBM-H2}, to give a top-down motivation for
a modified commutation relationship between position and momentum.  

\section{Modified Position and Momentum Commutator motivated by the 
Bender-Brody-M\"uller Hamiltonian}

We begin by writing down modified position and momentum operators in the form
\begin{equation}
\label{xp}
{\hat x'} = i (1 + g(p)) \partial_p ~~~~;~~~~ {\hat p'} = p (1+ f(p) ) ~.
\end{equation}

\noindent The expressions in \eqref{xp} represent
a general way of modifying the position and momentum operators. The form in \eqref{xp} includes the modified position and momentum operators from reference \cite{KMM-1994} if
one takes $f(p)=0$ and $g(p) = 1 + \beta p^2$. The form of the 
modified position and momentum operators in \eqref{xp} also covers 
the case of $\kappa$-deformed Poincar{\'e} algebra from reference
\cite{maggiore2}. In fact we will find that the form of the
$g(p)$ that we obtain in the end involves hyperbolic functions 
which gives a modified position operator similar in form to the modified Newton-Wigner position operator suggested in \cite{maggiore2}. Even more recently reference \cite{das} gave modified position and momentum
operators of the form in equation \eqref{xp} in order to formulate a relativistic generalized 
uncertainty principle.

To obtain a specific form for the generalized 
position and momentum operators in \eqref{xp} we require
that the new viral operator,  ${\hat x'} {\hat p'} 
+ {\hat p'} {\hat x'}$, with ${\hat x'}$ and  ${\hat p'}$  from \eqref{xp}, leads to a modified virial operator like that in \eqref{BBM-H2}. This requirement will lead to specific functions, $g(p)$ and $f(p)$, which in turn will give a specific form for the modified position and momentum operators. We are working in momentum space since the extra term in ${\hat H}$ from 
\eqref{BBM-H2} involves only the momentum operator. 
Using the operators from \eqref{xp} we find
that the new operator becomes
\begin{eqnarray}
\label{QG-H}
{\hat H} &=& ({\hat x'} {\hat p'} + {\hat p'} {\hat x'}) \\ 
&=& \left[ 2 i p 
(1+f(p))(1+ g(p)) \partial_p + i \right] 
+i (1+ g(p))(f(p) +p f'(p)) + i g(p) .\nonumber
\end{eqnarray}

\noindent The first term in \eqref{QG-H} 
({\it i.e.} $2 i p (1+ g(p))(1+f(p)) \partial_p + i$) 
should correspond to the first term in \eqref{BBM-H2} ({\it i.e.} 
$({\hat x} {\hat p} + {\hat p} {\hat x}) =2 i p 
\partial _p + i$, using ${\hat x} = i \partial _p$ and ${\hat p} 
= p$). This correspondence is accomplished by requiring $(1+f(p))(1+g(p)) = 1$, 
{\it i.e.} 
\begin{equation}
\label{gp}
g(p) = \frac{- f(p)}{1+f(p)} ~.
\end{equation}

\noindent With this $g(p)$ the remaining terms in \eqref{QG-H} become

$$
i (1+ g(p))(f(p) +p f'(p)) + i g(p) = \frac{i p f'(p)}{1+f(p)} ~.
$$

\noindent We determine $f(p)$ by requiring the above 
expression equal the last term in \eqref{BBM-H2} yielding

\begin{equation}
\label{fp}
\frac{i p f'(p)}{1+f(p)} = - \frac{2p \Delta x e^{-i p\Delta x}}{1- e^{-ip \Delta x}} \rightarrow 
\frac{d}{dp} \left[ \ln (1 +  f(p))\right] = \frac{2 i \Delta x e^{-ip \Delta x}}{1-e^{-ip \Delta}}~.
\end{equation}

\noindent Equation \eqref{fp} is straight forward to solve 
and yields the solution
\begin{equation}
\label{fp2}
1 + f(p) = C (1 - e^{-i p \Delta x} )^2 ~.
\end{equation}
\noindent from which it follows

\begin{equation}
\label{gp2}
1+g(p) = \frac{1}{C (1 - e^{-i p \Delta x} )^2  } ~.
\end{equation}

Using the modified position and momentum operators from equations
\eqref{xp} \eqref{fp2} and \eqref{gp2}, we find that the associated modified 
commutator becomes
\begin{equation}
\label{qg-cr}
[{\hat x'}, {\hat p'}] = i \left( 1 + \frac{p f'(p)}{1+ f(p)} \right) = 
i + \frac{2 p \Delta x}{1- e^{ip \Delta x}} ~,
\end{equation}

\noindent where we have used the expression for $g(p)$ from \eqref{gp} to 
get to the intermediate form, and to obtain the final form we used 
\eqref{fp2}. The first term, $i$, is the standard 
commutator, and the second term, $\frac{2 p \Delta x}
{1- e^{ip \Delta x}}$, is the modification coming from the deformation of the position and 
momentum operators. It is this second term which represents the
change that we associate with a modification of short distance/large
momentum behavior coming from quantum gravity.
Equation \eqref{qg-cr} is the modification of the quantum commutator
implied by the requirement that the modified position and momentum operators
from equations \eqref{xp}, \eqref{fp2}, and \eqref{gp2} give
the modified virial operator in \eqref{BBM-H} or \eqref{BBM-H2}.

We now impose the physical requirement that one should recover the standard 
operators in the low momentum limit, {\it i.e.} $g(p), f(p) \to 0$ as $p \to 0$.
It is easy to see from equations \eqref{fp2} and  
\eqref{gp2} that $f(p) \to -1$ and $g(p)$ diverges as $p \to 0$. Furthermore,  
as $p \to 0$, we want the commutator in equation \eqref{qg-cr} to go over
to $[{\hat x}, {\hat p}] = i$. As $p \Delta x \to 0$ we see that
$\frac{2p \Delta x}{1-e^{ip \Delta x}} \to 2i$, and thus in this limit the commutator
in \eqref{qg-cr} becomes $[{\hat x'}, {\hat p'}] \to 3i$ which is not
correct. As foreshadowed in the previous section, the operator from \eqref{BBM-H}, which is a 
combination of dilation and a difference operator, $1-e^{-i p \Delta x }$, does not modify the commutator 
in a way which leads to a minimal length scale.

Despite this initial failure we now ask if we can modify the 
operator in \eqref{BBM-H} or \eqref{BBM-H2} to get a modified 
commutator with the correct physical limit as $p \to 0$, while still
preserving the potential approach to the Riemann hypothesis proposed
in \cite{BBM-2017}.

We begin by finding a new ${\hat \Delta}$ which differs from the  
${\hat \Delta} _{BBM}$ from \eqref{BBM-H} and \eqref{BBM-H2} and which will give 
more physical behavior in the $p \to 0$ limit. We will also need to
check that this new ${\hat \Delta}$ still allows for the approach
to the Riemann Hypothesis given in \cite{BBM-2017}. One problem with the original construction is that
${\hat \Delta} _{BBM} =1 - e^{-ip \Delta x } \to 0$ as $p \Delta x \to 0$. To avoid this, we could
take a $+$ sign so ${\hat \Delta} = 1 + e^{-ip \Delta x }$. This new operator is 
a kind of averaging transformation of a function between points $x$ and
$x-\Delta x$ rather than a difference operator, as is the case
with ${\hat \Delta}_{BBM}$. Applying $1 + e^{-ip \Delta x}$
to a function, $f(x)$, one finds ${\hat \Delta} f(x)= f(x) + f(x-\Delta x)$. To make this a true averaging between the points $x$ and $x-\Delta x$, we should divide by $\frac{1}{2}$.

As mentioned in the introduction section we will consider a generalized averaging operator ${\hat \Delta_{ABS}} =  \frac{1}{2} (e^{kp} + e^{-kp})= \cosh (kp)$ given in \eqref{ABSDelta}. This ${\hat \Delta_{ABS}}$ 
is symmetric in $p$. Also unlike the Bender-Brody-M{\"u}ller case, when $k$ was imaginary 
({\it i.e.} $k= \pm i \Delta x$), here we find we need to have $k$ real. 
With ${\hat \Delta}_{ABS} =\cosh (kp)$, the modified virial operator becomes
\begin{eqnarray}
\label{H-cosh}
    {\hat H} &=& \hat{\Delta}_{ABS}^{-1}({\hat x}{\hat p} + {\hat p}{\hat x})\hat{\Delta}_{ABS}
    =  {\hat x}{\hat p} + {\hat p}{\hat x} + 2 p ({\hat \Delta}_{ABS}^{-1}[x,{\hat 
    \Delta}_{ABS}]) \nonumber \\
    &=& {\hat x}{\hat p} + {\hat p}{\hat x} + 
    \frac{2ip}{\cosh(kp)}\partial_p(\cosh(kp)) \\
    &=&  {\hat x}{\hat p} + {\hat p}{\hat x} + 2ipk\tanh(kp)~. \nonumber 
\end{eqnarray}
One obtains a differential equation, similar to \eqref{fp}, which 
reads $\frac{ipf'(p)}{1+f(p)} = 2ipk\tanh(kp)$. This gives the following solution 
for $f(p)$:
\begin{equation}
\label{fp1}
        1+f(p) = C\cosh^2(kp) ~,\\
\end{equation}
which can be used in \eqref{gp} to obtain $g(p)$
\begin{equation}
\label{gp1}
        1+g(p) = C^{-1}\text{sech}^2(kp).
\end{equation}

The functions $f(p) , g(p)$ from \eqref{fp1} and \eqref{gp1} inserted in 
\eqref{xp} give the modified position and momentum operators
\begin{equation}
\label{xp1}
    {\hat x}' = i\text{sech}^2(kp)\partial_p ~~~;~~~
    {\hat p}' = \cosh^2(kp)p ~
\end{equation}
where $C=1$ so ${\hat p}' \to p$ and ${\hat x}' \to i \partial _p$ as $p\to 0$. 
These operators are symmetric with respect to the inner product 
$\langle\psi(p)|\phi(p)\rangle = \int_{-\infty}^\infty \cosh^2(kp) 
\overline{\psi(p)}\phi(p)dp$.  This inner product leads to the norm $\|\psi\|^2 = 
\int_{-\infty}^\infty \cosh^2(kp) |\psi(p)|^2dp$. In order for this norm to be 
finite and give normalizable states, one needs exponential suppression of wave 
function at high momentum to counter the $\cosh^2(kp)$ factor. 

The modified commutation relations become
\begin{equation}
\label{xp-com}
    [{\hat x'} , {\hat p'}] = i\text{sech}^2(kp)\partial_p[\cosh^2(kp)p]\\
    = i\left(1 + 2kp\tanh(kp)\right)
\end{equation}
If $kp \ll 1$ the right hand side of \eqref{xp-com} can be expanded using $\tanh 
(kp) \approx kp + {\cal O} (kp)^3$ with the result 

\begin{equation}
\label{xp-com1}
    [{\hat x'} , {\hat p'}]  \approx i(1 + 2  k ^2 p^2).
\end{equation}
This $kp \ll 1$ limit gives a commutator which is the same as the 
phenomenological commutator given by \eqref{KMM-CR} with $\beta = 2 k^2$, provided that
$k$ is real so that $k^2 >0$. This is required since one needs $\beta >0$ in \eqref{KMM-CR}
in order to get a minimal length. For the operator ${\hat \Delta}_{ABS} = \cosh (kp)$, we
do require that $k$ be real and thus we recover a minimal length in a manner similar to  that in reference 
\cite{KMM-1994}. In contrast for the Bender-Brody-M{\"u}ller operator, ${\hat \Delta}_{BBM} = 1 - e^{-i p \Delta x}$,
one has $k= -i \Delta x$ which gives $k^2<0$ and no minimal length, which along with the results for
the $p \Delta x \to 0$ limit of the modified commutator from \eqref{qg-cr}, shows that the Bender-Brody-M{\"u}ller 
operator and Hamiltonian do not lead to a minimal length scale. This appearance of a minimal length scale,
for real $k$ ({\it i.e. } $k^2 >0$ and $\beta >0$), can be linked to the form of the modified virial operator  
in \eqref{H-cosh} which breaks the dilation symmetry associated with the virial
operator ${\hat x}{\hat p}+{\hat p} {\hat x}$, via the scale 
dependent averaging operator ${\hat \Delta}_{ABS} = \cosh (kp)$.

In the calculations leading to \eqref{xp-com} we have shown that by modifying 
the operator ${\hat \Delta}_{BBM}$ from $1- e^{-ip \Delta x}$ (the form taken in 
\cite{BBM-2017}) to $\cosh (kp)$ we get a physically reasonable modification 
of the position and momentum commutator.  One of the aims of this work was to tie 
the modification of the position and momentum commutator to
the Bender-Brody-M{\"u}ller variant of the Berry-Keating program. We will now
show that choosing ${\hat \Delta}_{ABS} = \cosh (kp)$ still allows one to follow a similar
construction to the one proposed in \cite{BBM-2017} to address the Riemann Hypothesis. 
We will find that there are several variants of ${\hat \Delta}$ which work.

The basic idea of the Berry-Keating program is that there exists some operator, 
${\hat H}$, which satisfies an eigenvalue equation ${\hat H} \Psi = E \Psi$ 
whose eigenvalues, $E$, give the imaginary part of the non-trivial
zeros of the Riemann zeta function. The operator we consider is of the 
Bender-Brody-M{\"u}ller form ${\hat H} = {\hat \Delta}^{-1} ({\hat x} 
{\hat p} + {\hat p} {\hat x}) {\hat \Delta}$ where we take
${\hat \Delta}_{ABS} = \cosh (kp)$. Following \cite{BBM-2017} we begin by 
re-writing the eigenvalue equation as 
\begin{equation}
\label{EV-1}
{\hat H} \Psi = E \Psi \to ({\hat x} {\hat p} + {\hat p} {\hat x}) ({\hat \Delta}_{ABS} 
\Psi) = E ({\hat \Delta}_{ABS} \Psi)~,
\end{equation}
which is an eigenvalue equation for ${\hat \Delta}_{ABS} \Psi$ with respect to the 
operator ${\hat x} {\hat p} + {\hat p} {\hat x}$. Using the standard coordinate 
space representation \footnote{Here we switch from momentum 
space operators to coordinate space operators to follow the construction
of reference \cite{BBM-2017}.} of the position and 
momentum operators, ${\hat x} = x$ and ${\hat p} = - i \partial _x$,  the 
eigenfunctions and eigenvalues to \eqref{EV-1} are ${\hat \Delta}_{ABS} \Psi (z , x) = A 
x^{-z}$ and $E_z=i(2 z -1)$ respectively, where $A$ is a constant. 

We want to find $\Psi$ such that ${\hat \Delta}_{ABS} \Psi  (z , x)  = A x^{-z}$. If we apply 
$e^{kp}$ to an analytic function, $f(x)$ the result is  $e^{kp} f(x) = f ( x - i k )$ 
{\it i.e.} $e^{kp}$ is a generalized shift operator which shifts $f(x)$ by $ik$. 
When $k= \pm i$ this is a shift of $x \to x \pm 1$. It follows $\cosh(kp) f(x) = \frac{1}{2} (e^{kp} + 
e^{-kp}) f(x) = \frac{1}{2} [f ( x - i k ) + f ( x + ik )]$.  
In reference \cite{BBM-2017} where ${\hat \Delta} = 1 - e^{-ip}$, the
solution to ${\hat \Delta} \Psi  (z , x) = A x^{-z}$ was the Hurwitz zeta 
function, $\zeta(z, x+1)$, defined as
\begin{equation}
\label{HZ}
\Psi  (z, x) \propto \zeta (z, x+1) = \sum _{n=0} ^\infty \frac{1}{(n+x+1)^z} ~.
\end{equation}
By imposing the boundary condition $\Psi (z, 0) = 0$, this `forces' the Riemann 
zeta function $\zeta (z, 1)$ to equal $0$.
If the spectrum of this operator can be shown to be real, then the $z$'s have the 
form $\frac{1}{2} + it$ for real $t$, {\it i.e.} 
the non-trivial zeros of the zeta function are on this critical line.  
We have no new argument on the reality of the spectrum.  

For the case when  ${\hat \Delta}_{ABS} = \cosh (kp) = 
\frac{1}{2} ( e^{kp} + e^{-kp})$,  we use the Hurwitz-Euler eta function \cite{Choi2011}
\begin{equation}
\label{heta}
\eta(z,x+1) = \sum\limits_{n=0}^\infty \frac{(-1)^{n}}{(n+x+1)^z}  ~,
\end{equation}
and show that it solves the equation ${\hat \Delta}_{ABS} \Psi  (z , x)  = A x^{-z}$. 
This function is an alternating sign version of the Hurwitz zeta function of 
\eqref{HZ}. For $x=0$, $\eta(z,x+1)$ becomes the well known Dirichlet eta function 
\cite{stegun}
\begin{equation}
   \label{deta}
\eta(z,1) = \sum\limits_{n=0}^\infty \frac{(-1)^{n}}{(n+1)^z} ~,
\end{equation}
an alternating sign version of the Riemann zeta function. 
 
The Hurwitz-Euler eta function  satisfies ${\hat \Delta}_{ABS} \Psi  (z , x)  = A x^{-z}$
by applying the shift $x \to \frac{x}{2ik} - \frac{1}{2}$ in \eqref{heta} and 
arrive \footnote{so $x+1 \to \frac{x}{2ik} + \frac{1}{2}$} at
\begin{equation}
\label{heta-1}
\eta \left( z,\frac{x}{2ik} + \frac{1}{2} \right) = \sum\limits_{n=0}^\infty 
\frac{(-1)^{n}}{(n + \frac{x}{2ik} + \frac{1}{2})^z} ~.
\end{equation}
Recalling that ${\hat \Delta}_{ABS} f(x) = \frac{1}{2} (f(x-ik) + f(x+ik))$ and applying 
this to $\eta \left( z,\frac{x}{2ik} + \frac{1}{2} \right)$ yields
 \begin{eqnarray}
 \label{EV-2}
    \hat{\Delta}_{ABS} \left[ \eta \left( z,\frac{x}{2ik} + \frac{1}{2} \right) \right] &=& \frac{1}{2} 
    \left[ \eta \left( z, \frac{x}{2ik} \right) + \eta \left( z, \frac{x}{2ik} +1 
    \right) \right] \nonumber \\
    &=& 
\frac{1}{2}  \sum\limits_{n=0}^\infty \frac{(-1)^{n}}{(n + \frac{x}{2ik} )^z}
 +  \frac{1}{2} \sum\limits_{n=0}^\infty \frac{(-1)^{n}}{(n + \frac{x}{2ik} +1)^z} 
 \\
 &=&   \frac{1}{2}  \left(\frac{x}{2ik} \right)^{-z} \propto x^{-z} \nonumber ~,
 \end{eqnarray}
where the alternating sign of the two series makes all the terms cancel between 
the series except for the $n=0$ term of the first series. This shows that 
$\eta \left( z,\frac{x}{2ik} + \frac{1}{2} \right)$ satisfies the equation 
${\hat \Delta}_{ABS} \Psi  (z , x)  = A x^{-z}$. 

To continue the Bender-Brody-M{\"u}ller approach, we impose the boundary condition 
that the functions should be equal to zero at $x=ik$, {\it i.e.}
$\eta \left( z,\frac{ik}{2ik} + \frac{1}{2} \right) = \eta (z, 1) = 0$, which 
makes the Dirichlet eta function equal to zero. The Dirichlet eta function has the 
same non-trivial zeros as the Riemann zeta function. This can be seen through the 
functional relationship between the Riemann zeta function
and Dirichlet eta function \cite{stegun}:
$$\eta(z, 1) = (1-2^{1-z})\zeta(z, 1) .$$
Thus both the trivial and non-trivial zeros of the Riemann zeta function are zeros 
of the Dirichlet eta 
function.  The Dirichlet eta  function has additional trivial zeros of the form 
$z= 1 + 2\pi i k/\ln(2)$ with $k\in \mathbb{Z}$ so the pre-factor $(1-2^{1-z}) = 
0$. In \cite{BBM-2017}, it is argued that 
the trivial zeros at $z=-2n$ of $\zeta(z, 1)$ correspond to the eigenfunctions
which diverge as $x \to \infty$ and thus do not belong to the function space. 
The additional trivial zeros for the Dirichlet eta function at $z= 1 + 2\pi i 
k/\ln(2)$ could also be discarded for similar function space reasons. However, we note 
that without a well-defined function space, these arguments are suggestive at best.
The boundary condition creates a correspondence between the non-trivial zeros of 
the Dirichlet eta function and the solutions of the eigenvalue equation.
These zeros are exactly the non-trivial zeros of the Riemann zeta function and 
this shows that the approach in \cite{BBM-2017} to addressing the Riemann Hypothesis
also works for ${\hat \Delta}_{ABS} = \cosh (kp)$.

One difference of the present construction versus that the Bender-Brody-M{\"u}ller 
(aside from the use of Hurwitz-Euler eta and Dirichlet eta functions versus 
Hurwitz zeta and Riemann zeta functions) is that now the non-trivial zeros 
are determined by setting the function equal to zero at $x=ik$ as opposed
to $x=0$ as in \cite{BBM-2017}. The reason for this
shift of the location of the zeros of the eigenfunction, from $x=0$ to $x=ik$, can 
be seen by considering ${\hat \Delta} = \frac{1}{2} (1 + e^{2kp})$. One can follow 
through the steps in equations \eqref{EV-1} - \eqref{EV-2} and show that ${\hat 
\Delta}= \frac{1}{2} (1 + e^{2kp})$ also works 
for a construction similar to that given by Bender-Brody-M{\"u}ller. 
The $\Psi$ satisfying  ${\hat \Delta} \Psi  (z , x)  = A x^{-z}$ is now of the 
form $\Psi(z,x) = \eta (z, \frac{x}{2ik})$ (note the lack of $+ \frac{1}{2}$).
Thus for ${\hat \Delta} = \frac{1}{2} (1 + e^{2kp})$, the boundary condition 
is set at $x=0$, {\it i.e.} $\eta (z, x=0) = 0$. 
Finally, we can get from ${\hat \Delta} = \frac{1}{2} (1 + e^{2kp})$
to ${\hat \Delta}_{ABS} = \cosh (kp) =  \frac{1}{2} ( e^{kp} + e^{-kp})$ by applying 
$e^{-kp}$ to $\frac{1}{2} (1 + e^{2kp})$. The operator $e^{-kp}$ 
shifts functions by $ik$ ({\it i.e.} $e^{-kp} f(x) \to f(x+ik)$) which would shift
the boundary condition from $x=0$, for the ${\hat \Delta} =  \frac{1}{2} (1 + 
e^{2kp})$ case, to $x=ik$, for the ${\hat \Delta}_{ABS} = \cosh (kp) =  \frac{1}{2} ( 
e^{kp} + e^{-kp})$ case.

\section{Summary and Remarks}

The main result of this work is that we arrive at a modification
of the standard quantum position and momentum commutation relationship, using the
Bender-Brody-M{\"u}ller variant of the Berry-Keating program as a
guide to give a specific form for the modified commutator. 
These modified operators and commutators are given in equations 
\eqref{xp1} and \eqref{xp-com}. This differs from earlier proposals for modified 
operators and commutators, such as \eqref{KMM-CR}, which are phenomenologically motivated. 
The modified operators and commutators lead to a minimum length scale and a modified dilation symmetry generated by ${\hat \Delta}_{ABS}^{-1} ({\hat x} 
{\hat p} + {\hat p} {\hat x}) {\hat \Delta}_{ABS}$.  
In addition to providing a 
theoretical, top-down approach to writing down the modified commutators,
we also found that several different variants of the ${\hat \Delta}$ used
in defining the operator ${\hat H} = {\hat \Delta}^{-1} ({\hat x} 
{\hat p} + {\hat p} {\hat x}) {\hat \Delta}$ allow one to tackle
the Riemann hypothesis in the way proposed in reference \cite{BBM-2017}. In 
addition to ${\hat \Delta_{BBM}} = 1-e^{-i {\hat p \Delta x}}$, used by
Bender-Brody-M{\"u}ller, we have found that ${\hat \Delta}_{ABS} = \frac{1}{2} 
(e^{kp}+e^{-kp})$ and ${\hat \Delta} = \frac{1}{2}  (1+e^{2kp})$ also
lead to similar approaches to the Riemann hypothesis. 

We conclude with the remark that the analysis here can be
used to show that modifications of the quantum commutation, such as given in 
\eqref{KMM-CR}, can be connected with different modifications of the position and 
momentum operators.  In \cite{KMM-1994} the modified position and momentum 
operators connected with $[\hat{x},\hat{p}] = i(1+\beta \hat{p}^2)$ were given as
\begin{equation}
\label{xp-kmm}
{\hat x} = i (1 + \beta p^2 ) \partial _p ~~~;~~~
{\hat p} = p ~.
\end{equation}
The position operator is changed but the momentum operator is
not. Using the analysis of position and momentum operators
starting with \eqref{xp}, but having in mind the modified
commutator given in \eqref{KMM-CR}, we find that the ansatz functions are 
$1+f(p)=e^{\beta p^2/2}$ and $1+g(p) = e^{-\beta p^2/2}$.  These lead to modified 
position and momentum operators of the form
\begin{equation}
\label{xp-kmm1}
{\hat x}'= ie^{-\beta p^2/2}\partial_p ~~~;~~~ {\hat p}'=e^{\beta p^2/2}p ~.
\end{equation}
Both sets of modified operators -- those from equation \eqref{xp-kmm} and equation 
\eqref{xp-kmm1} -- lead to the same modified commutation relationship 
\eqref{KMM-CR}.

\section*{Acknowledgement} The authors would like to acknowledge the role of the FAMP (Functional
Analysis and Mathematical Physics) seminar in introducing them to this problem and leading to 
the discussions that resulted in this work.

\end{document}